\documentclass[prx,twocolumn,floatfix,superscriptaddress,longbibliography]{revtex4-1}

\usepackage{amssymb,amsmath,amstext,amsthm,mathtools}
\usepackage{graphicx}
\usepackage{epstopdf}
\usepackage{color}
\usepackage{bm}
\usepackage{appendix}
\usepackage[T1]{fontenc}
\usepackage{bbold}
\usepackage{bbm}
\usepackage{latexsym}
\usepackage[colorlinks=true,citecolor=blue,linkcolor=red,urlcolor=black]{hyperref}
\usepackage{float}
\usepackage{verbatim}
\usepackage{longtable}
\usepackage{fullpage}
\usepackage{makecell}
\usepackage{tabularx}
\usepackage{array}
\usepackage{comment}
\usepackage{lipsum}
\usepackage{color}
\usepackage{pgf}
\usepackage{tabularx}
\usepackage{array, makecell} %
\newcolumntype{L}{>{\raggedright\arraybackslash}X}
\newcolumntype{P}[1]{>{\centering\arraybackslash}p{#1}}

\newcommand{\ket}[1]{| #1 \rangle}

\newcommand{\beq}{\begin{equation}}
\newcommand{\eeq}{\end{equation}}

\newcommand{\mte}[2]{\langle#1|#2|#1\rangle }

\DeclareMathOperator*{\argmax}{arg\,max}
\DeclareMathOperator*{\argmin}{arg\,min}
\renewcommand{\vec}[1]{\boldsymbol{#1}}  

\newcommand*{\id}{\openone}

\newtheoremstyle{example}{\topsep}{\topsep}%
{}
{}
{\bfseries}
{:}
{   }
{\thmname{#1}\thmnumber{ #2}}
\theoremstyle{example}

\theoremstyle{definition}

\begin{document}

\title{Quantum autoencoders with enhanced data encoding}

\author{Carlos Bravo-Prieto}
\affiliation{Departament de F\'isica Qu\`antica i Astrof\'isica and Institut de Ci\`encies del Cosmos (ICCUB), Universitat de Barcelona, Mart\'i i Franqu\`es 1, 08028 Barcelona, Spain.}
\affiliation{Quantum Research Centre, Technology Innovation Institute, Abu Dhabi, UAE.}

\begin{abstract}
We present the enhanced feature quantum autoencoder, or EF-QAE, a variational quantum algorithm capable of compressing quantum states of different models with higher fidelity. The key idea of the algorithm is to define a parameterized quantum circuit that depends upon adjustable parameters and a feature vector that characterizes such a model. We assess the validity of the method in simulations by compressing ground states of the Ising model and classical handwritten digits. The results show that EF-QAE improves the performance compared to the standard quantum autoencoder using the same amount of quantum resources, but at the expense of additional classical optimization. Therefore, EF-QAE makes the task of compressing quantum information better suited to be implemented in near-term quantum devices.

\end{abstract}

\maketitle

\section{Introduction} \label{sec:introduction}
Large-scale fault-tolerant quantum computation is a rather distant dream, typically estimated to be a few decades ahead. A reasonable question then is whether we can do something useful with the existing noisy intermediate-scale quantum (NISQ)~\cite{preskill2018quantum, bharti2021noisy} computers. The main proposal is to use them as a part of a hybrid classical-quantum device. The variational quantum algorithms (VQAs)~\cite{cerezo2020variational} are a class of algorithms that use such hybrid devices, which manage to reduce the requisites of quantum computational resources at the expense of classical computation.

The general rationale of a VQA is to define a parametrized quantum circuit whose architecture is dictated by the type and size of the quantum computer that is available. This quantum circuit, in turn, will depend on a set of classical parameters that can be adjusted using a quantum-classical optimization loop by minimizing a cost function. In this manner, we look for a quantum circuit that allows to perform a particular task, given the available quantum resources. Let us remark here that several VQAs have already been proposed in the context of making NISQ computers practically useful for real applications ~\cite{VQE, Kokail, VQEex, Jones, Li, QAE, QAQC, VQSD, VQSVD, vqls, holmes2019sim, carolan2019, mcardle2019evolution, endo2020variational, uvarov2020variational, borzenkova2021variational}.

Recently, much attention has been paid to data encoding in VQAs  \cite{lloyd2020quantum, larose2020robust}, since it was proven that data encoded into the model alters the expressive power of parameterized quantum circuits \cite{schuld2020effect, goto2020universal}. Specifically, this idea has been implemented for classification of data \cite{havlivcek2019supervised, perez2020data}, and to study energy profiles of quantum Hamiltonians \cite{cervera2020meta}.

\begin{figure}[t]
    \centering
    \includegraphics[width=1.0\linewidth]{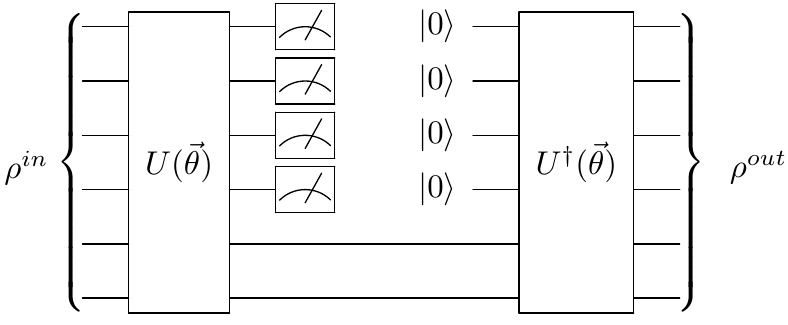}
	\caption{Circuit implementation of a quantum autoencoder with a 2-qubit latent space. The unitary $U(\vec{\theta})$ encodes a 6-qubit input state $\rho^{in}$ into a 2-qubit intermediate state, after which the decoder $U^\dagger(\vec{\theta})$ attempts to reconstruct the input, resulting in the output state $\rho^{out}$.}
	\label{fig:autoencoder}
\end{figure}

In this paper, we will explore how data encoding influences a Quantum Autoencoder (QAE) \cite{QAE}. The QAE is a VQA designed to compress the input quantum information through a smaller latent space. In this scheme, we look for a parameterized quantum circuit $U(\vec{\theta})$ that encodes an initial input state into an intermediate latent space, after which the action of the decoder, $U^\dagger(\vec{\theta})$, attempts to reconstruct the input. A graphical depiction of a QAE is shown in Fig.~\ref{fig:autoencoder}. For readers interested in experimental applications, a QAE implementation in a photonic device can be seen in Ref.  \cite{pepper2019experimental}.

Note that the motivation for a quantum autoencoder is to be able to recognize patterns beyond the capabilities of a classical autoencoder, given the different properties of quantum mechanics. Moreover, recall that for NISQ devices, any tool that can reduce the amount of quantum resources can be considered valuable. For instance, quantum autoencoders could be used as a state preparation engine in the context of other VQAs. That is, we could combine, say, a Variational Quantum Eigensolver \cite{VQE} with a pretrained QAE, where now the only active parameters are associated with the latent space.

\begin{figure*}[t!]
	\centering
	\includegraphics[width=1.0\linewidth]{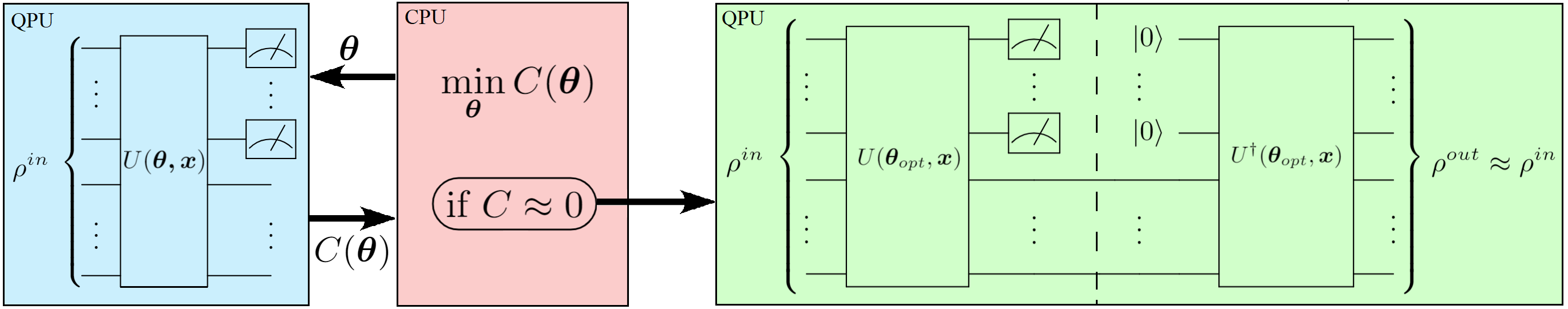}
	\caption{Schematic representation of the EF-QAE. The input to EF-QAE is a set of initial states $\rho^{in}$, a feature vector $\vec{x}$ that characterizes the initial states, and a shallow sequence of quantum gates $U$.  The feature vector $\vec{x}$ is encoded together with the variational parameters $\vec{\theta}$, where the latter are adjusted in a quantum-classical optimization loop until the local cost $C(\vec{\theta})$ converges to a value close to 0. When this loop terminates and the optimal parameters $\vec{\theta}_{opt}$ are found, the resulting circuit $U(\vec{\theta}_{opt}, \vec{x})$ prepares compressed states $\ket{\phi}$ of a particular model. Moreover, we may apply $U^\dagger(\vec{\theta}_{opt}, \vec{x})\ket{0\dots0}\otimes\ket{\phi}$ to recover $\rho^{out} \approx \rho^{in}$.} 
	\label{fig:EF-QAE-diagram}
\end{figure*}

This paper is organized as follows. In Sec. \ref{sec:algorithm} we introduce the enhanced feature quantum autoencoder (EF-QAE). As we will see, its key ingredient is to include a feature vector into the variational quantum circuit that characterizes the model we aim to compress. Next, in Sec. \ref{sec:ising} and Sec. \ref{sec:handwritten} we compare and assess the performance of the EF-QAE and the standard QAE in simulations, by compressing ground states of the 1D Ising model and classical handwritten digits, respectively. Finally, in Sec. \ref{sec:discussion}, we present the conclusions of this work.

\section{EF-QAE Algorithm} \label{sec:algorithm}

\subsection{Overview} \label{sec:overview}
Here, we present the enhanced feature quantum autoencoder (EF-QAE). A schematic diagram of the EF-QAE can be seen in Fig. \ref{fig:EF-QAE-diagram}. The algorithm can be initialized with a set of initial states $\rho^{in}_i$, a feature vector $\vec{x}$, and a shallow sequence of quantum gates $U$. In this scheme, we define a unitary $U(\vec{\theta, x})$ acting on the initial state $\rho^{in}_i$, where $\vec{x}$ is a feature vector that characterizes the set of input states. For instance, as we will see in Sec. \ref{sec:ising}, $\vec{x}$ may be the transverse field $\lambda$ of the 1D Ising spin chain. Once the trial state is prepared, measurements are performed to evaluate the cost function $C(\vec{\theta})$. This result is then fed into the classical optimizer, where the parameters $\vec{\theta}$ are adjusted. This quantum-classical loop is repeated until the cost function converges to a value close to 0. When the loop terminates, $U(\vec{\theta}_{opt}, \vec{x})$ prepares compressed states $\ket{\phi}$ of a particular model.

A summary comparing EF-QAE and QAE proposed in Ref. \cite{QAE} can be seen in Appendix \ref{sec:comparison}. Note that the main difference between EF-QAE and QAE is the presence of a feature vector $\vec{x}$ in the sequence of gates $U$. This will allow us to study and explore how data encoding influences the behavior of a quantum autoencoder.

\subsection{Cost function} \label{sec:costfunction}
The goal of a quantum autoencoder is to store the quantum information of the input state through the smaller latent space. Therefore, it is important to quantify how well the information is preserved. This in general is quantified by a cost function that one has to minimize. In Ref. \cite{QAE}, this cost function evaluates the fidelity of the input and output states, and it is constructed from global operators. However, it is known that global cost functions lead to trainability issues even for shallow depth quantum circuits \cite{Mcclean2018barren, cerezo2020cost}. 

To address this issue, we use a cost function designed from local operators, proposed in Ref. \cite{cerezo2020cost}. As mentioned therein, there is a close connection between data compression and decoupling. That is, if the discarded qubits, from now on referred to as trash qubits, can be perfectly decoupled from the rest, the autoencoder reaches lossless compression. For instance, if the output of the trash subsystem is a fixed pure state, say $\ket{0...0}$, then it is decoupled and consequently, the input state has been successfully compressed.

A figure of merit to quantify the degree of decoupling, or data compression, when training is simply the total amount of non-zero measurement outcomes on the $n_t$ trash qubits, which will be minimized. To design the cost function to be local, different outcomes may be penalized by their Hamming distance to the $\ket{0}^{\otimes n_t}$ state, which is just the number of symbols that are different in the binary representation.  Thus, the local cost function $C$ to be minimized is
\beq \label{eq:cost_function} C\equiv \sum_{k,j} d_{Hj} M_{k,j} \equiv \frac{1}{2} \sum_{k=1}^{n_t} (1 - \langle Z_k\rangle)\,,\eeq
where $d_{Hj}$ denotes the Hamming distance and $M_{k,j}$ are the results of the $j$-th measurement on the $k$ trash qubit in the computational basis. Equivalently, it can also be defined in terms of local $Z$ Pauli operators. Finally, notice that this cost function delivers direct information on how the compression of the trash qubits is performed and has a zero value if and only if the compression is completed.

\subsection{Ansatz} \label{sec:ansatz}
To implement the EF-QAE model on a quantum computer, we must define the form of the parametrized unitary $U(\vec{\theta, x})$, decomposing it into a quantum circuit suitable for optimization. Recall that a quantum autoencoder may be thought of as a disentangling unitary. The complexity of the circuit thus limits this property. Given the limited available quantum resources in practice, due to the coherence times and gate errors, we will look for a circuit structure that maximally exploits entanglement while maintaining a shallow depth.

\begin{figure}[t]
    \centering
    \includegraphics[width=\columnwidth]{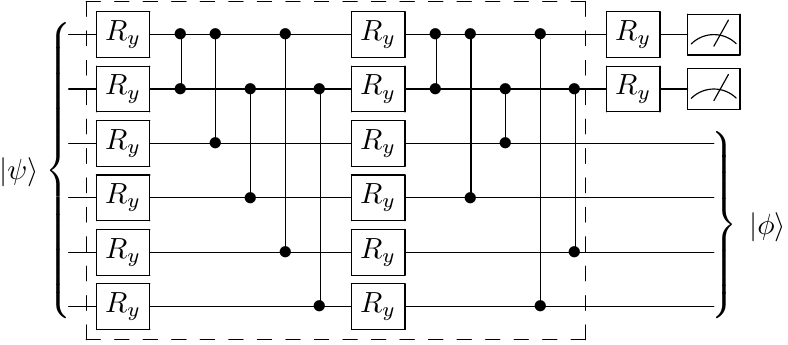}
    \caption{Variational quantum ansatz employed for the EF-QAE model. As indicated by the dashed box, each layer is composed of $CZ$ gates acting on the trash qubits preceded by $R_y$ qubit rotations, $R_y (\theta_j)= e^{-i\theta_j Y/2}$. A cascade of $CZ$ gates is then applied between the trash qubits and the qubits containing the final compressed state. After implementing the layered ansatz, a final layer of $R_y$ qubit gates is applied to the trash qubits. Note that the sequence of entangling gates can be applied mostly in parallel.}
    \label{fig:new_ansatz}
\end{figure}

A primitive strategy to construct a variational circuit in a more general case may consist of building a circuit of arbitrary 2- and 1-qubit gates characterized by some parameters. However, this is a naive approach. The action of the EF-QAE on the original state is
\beq \label{eq:autoencoder} U \ket{\psi} =  \ket{0} \otimes \hdots \otimes \ket{0} \otimes \ket{\phi}\,.\eeq
Thus, it is clear that the entangling gates should mostly act between each of the trash qubits, and between the trash qubits and the qubits containing the final compressed state. Subsequently, we may avoid using entangling gates between the qubits that are not trash while maximizing the entangling gates on the ones of interest. This could be done using a similar structure to that depicted in Fig. \ref{fig:new_ansatz}. Notice that most of the sequence of entangling gates can be applied in parallel at the same step, and that the number of quantum gates is linear with the number of qubits and layers.

In this work, we follow a similar encoding strategy to that in Ref. \cite{perez2020data}. That is, we encode the feature vector $\vec{x}$ into each of the single $R_y$ qubit rotations by using a linear function as
\beq \label{eq:ry} R_y^{(i,j)}\left(\vec{\theta, x}\right) = R_y\left(\theta^{(i)} x^{(j)} + \theta^{(i+1)}\right)\,,\eeq
where $i,j$ indicates a component of the vector, and $\vec{\theta}$ are the parameters adjusted in the optimization loop. 

The rationale behind choosing this kind of encoding is that it has been shown to provide universality, provided enough layers, and with a single qubit \cite{perez2020data}. Here, although we use multiple qubits, and entanglement is allowed, we expect a similar behavior as the number of layers increases. Note as well that this encoding is clearly analogous to that used in classical neural networks. That is, $\vec{\theta}$ plays the role of the weights and biases, while the rotation gate plays the role of the non-linear activation function. On the other hand, the role of the feature vector $\vec{x}$ is inspired by feed-forward classical neuronal networks. Specifically, in this type of classical network, data is reintroduced and processed by many layers of neurons, similar to what our quantum circuit is doing. From a quantum mechanical perspective, we can say that the quantum data compression is tailored to a particular input, informed by the feature vector $\vec{x}$. That is, EF-QAE is applying different unitary operations $U(\vec{\theta}, \vec{x})$ to different input states, depending on the extra information delivered by the feature vector $\vec{x}$, and by doing so, improving the compression performance.

Lastly, let us remark that other encoding strategies of the feature vector can be considered, for instance, using a non-linear encoding \cite{cervera2020meta}.

\section{1D Ising spin chain} \label{sec:ising}
The EF-QAE can be verified on simulations. We utilized the open-source Python API {\tt Qibo} \cite{qibo, efthymiou2020qibo} for the simulation of the quantum circuits. Here, we benchmark both the EF-QAE and the standard QAE in the case of a paradigmatic quantum spin chain with 6 qubits, the transverse field Ising model. The 1D Ising model is described by the following Hamiltonian 
 \begin{equation} \label{eq:ising}
 H_{Ising} = \sum_j \sigma_j^z \sigma_{j+1}^z + \lambda \sum_j \sigma_j^x\,,
 \end{equation}
where $\lambda$ is the transverse field. In the thermodynamic limit, the system has a quantum phase transition exactly at $\lambda=1$.

\begin{figure}[t]
    \centering
    \includegraphics[width=\columnwidth]{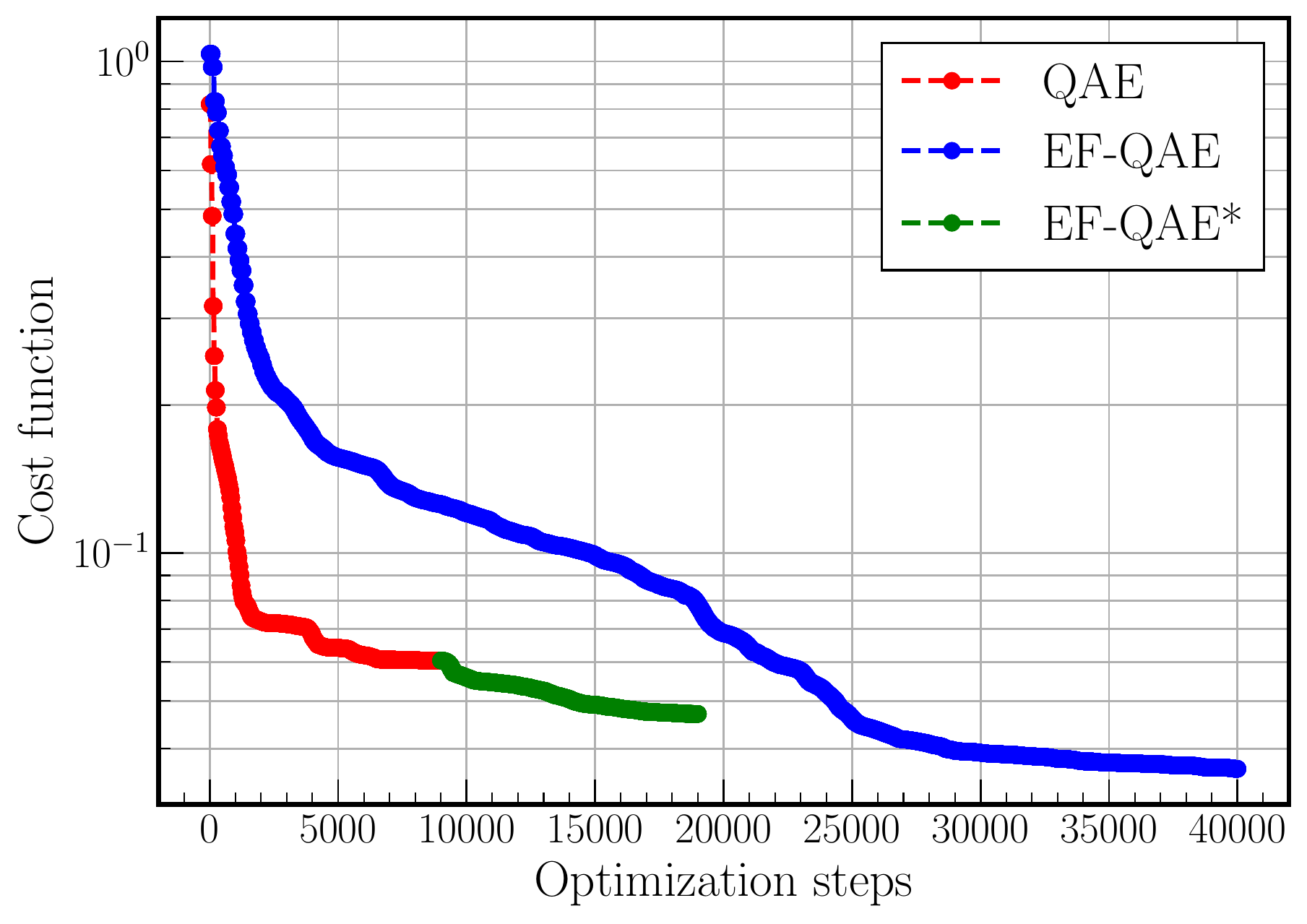}
    \caption{Cost function value as a function of the number of evaluations. Here, we consider the standard QAE, EF-QAE, and EF-QAE*. The EF-QAE* is the EF-QAE initialized with the optimal parameters of QAE. The EF-QAE achieves almost twice the compression of the QAE using the same quantum resources, at the expense of additional classical optimization.}
    \label{fig:ising-steps}
\end{figure}

The EF-QAE and QAE are optimized over a training set of ground states of the Ising model. Specifically, we have considered N=20 equispaced ground states in between $\lambda=0.5$ and $\lambda=1.0$, with initial random parameters. For the cost function, we computed Eq.~\ref{eq:cost_function} for each training state and then averaged them as
\begin{equation} \label{eq:cost}
C_N = \frac{\sum^N_i C_i}{N}\,.
 \end{equation}
Nonetheless, notice that for other models, sophisticated cost functions could be more convenient to implement. We have considered the variational quantum circuit in Fig.~\ref{fig:new_ansatz} with 3 layers, and therefore, the resulting compressed state contains 4 qubits. Here, the feature vector $\vec{x}$ for the EF-QAE is a scalar that takes the value of the transverse field $\lambda$. 

The classical technique employed in the optimization loop is the BFGS method, which is gradient-based and involves estimation of the inverse Hessian matrix \cite{nocedal2006numerical}. Let us also briefly comment here on the training required for both QAE and EF-QAE. Indeed, although the depth of the circuit is equivalent, the number of trainable parameters is not. In this sense, QAE has $1$ trainable parameter on each rotation-gate, whereas EF-QAE has dim($\vec{x}$) + 1 trainable parameters. For this example, dim($\vec{x}$) $= 1$, since $\vec{x}$ is just a scalar value, and therefore, the number of trainable parameters is 2. For gradient-based optimizers, this may imply the computation of extra gradients, and therefore, extra cost function evaluations. Recall, however, that this possible classical overhead is only present during the training procedure, and hence, we will not face any overhead when using a pretrained EF-QAE in combination with other machine learning tasks.

\begin{figure}[t!]
    \centering
    \includegraphics[width=\columnwidth]{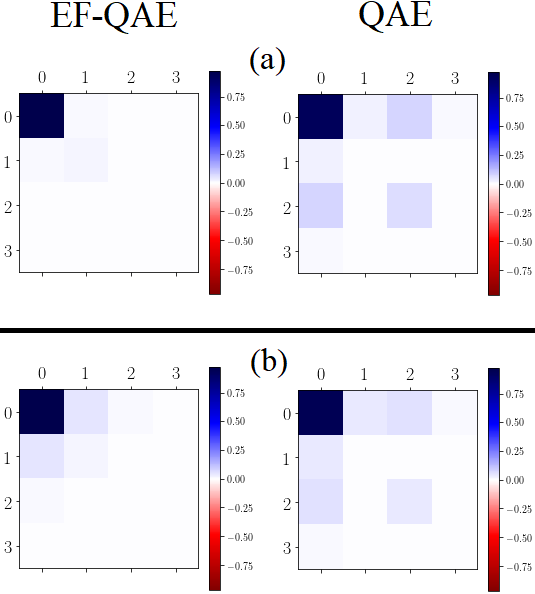}
    \caption{Visualization of the trash space for the EF-QAE and QAE, considering two different test ground states of the 1D Ising model corresponding to (a) $\lambda=0.60$ and (b) $\lambda=0.75$. The size of the register is two qubits. The space is characterized as the density matrix of the trash state. Integer labels denote the binary representation of the computational basis states.}
    \label{fig:ising-space}
\end{figure}

In Fig. \ref{fig:ising-steps}, we show the cost function value as a function of the number of evaluations. The EF-QAE* is the EF-QAE initialized with the optimal parameters of QAE. This way, the EF-QAE* will always improve the QAE performance. As can be seen, the EF-QAE achieves almost twice the compression of the QAE. Nevertheless, notice that for the EF-QAE, the number of function evaluations required to achieve higher compression is larger. Recall that this is simply a trade-off between classical and quantum resources. That is, using the same quantum resources we improve the compression performance at the expense of additional classical optimization.

To quantify these expectations, we assess both EF-QAE and QAE with the optimal parameters against two test ground states of the Ising model, specifically, with $\lambda=0.60$ and $\lambda=0.75$. The results are shown in Fig.~\ref{fig:ising-space}. Here, we show a density matrix visualization of the trash space. The EF-QAE achieves better compression to the $\ket{00}$ trash state, and therefore, higher fidelity on the output state. As we change the values of the transverse field, we note however that compression differs. In Appendix \ref{sec:fidelities} we discuss and provide the output fidelities of the training and 60 test ground states.

\section{Handwritten digits} \label{sec:handwritten}
In this section, we benchmark EF-QAE and QAE models in the case of $8\times8$ handwritten digit compression with 6 qubits using 4 layers. The data comprising each digit consists of a matrix with values from 0 to 16 corresponding to a gray map. Each value of this matrix is encoded in the amplitude of a 6-qubit state, further restricted to normalization.

The EF-QAE and QAE are optimized over a training set of handwritten digits obtained from the Python package {\tt Scikit Learn}~\cite{scikit-learn}. Specifically, we have considered N=20 handwritten digits, 10 of each corresponding to $\mathbf{0}$ or $\mathbf{1}$. The simulation details are equivalent to those in Sec.~\ref{sec:ising}. Here, the feature vector for the EF-QAE corresponds to $x=1, 2$. That is, we simply input a value of $x=1$ $(x=2)$ if the handwritten digit corresponds to $\mathbf{0}$ $(\mathbf{1})$. The reason to choose $x=1, 2$ is that no obvious feature distinguishes both digits. Nonetheless, more convenient strategies could be used in future work. For instance, one may allow the feature vector $\vec{x}$ to be a free variational parameter.

\begin{figure}[t]
    \centering
    \includegraphics[width=\columnwidth]{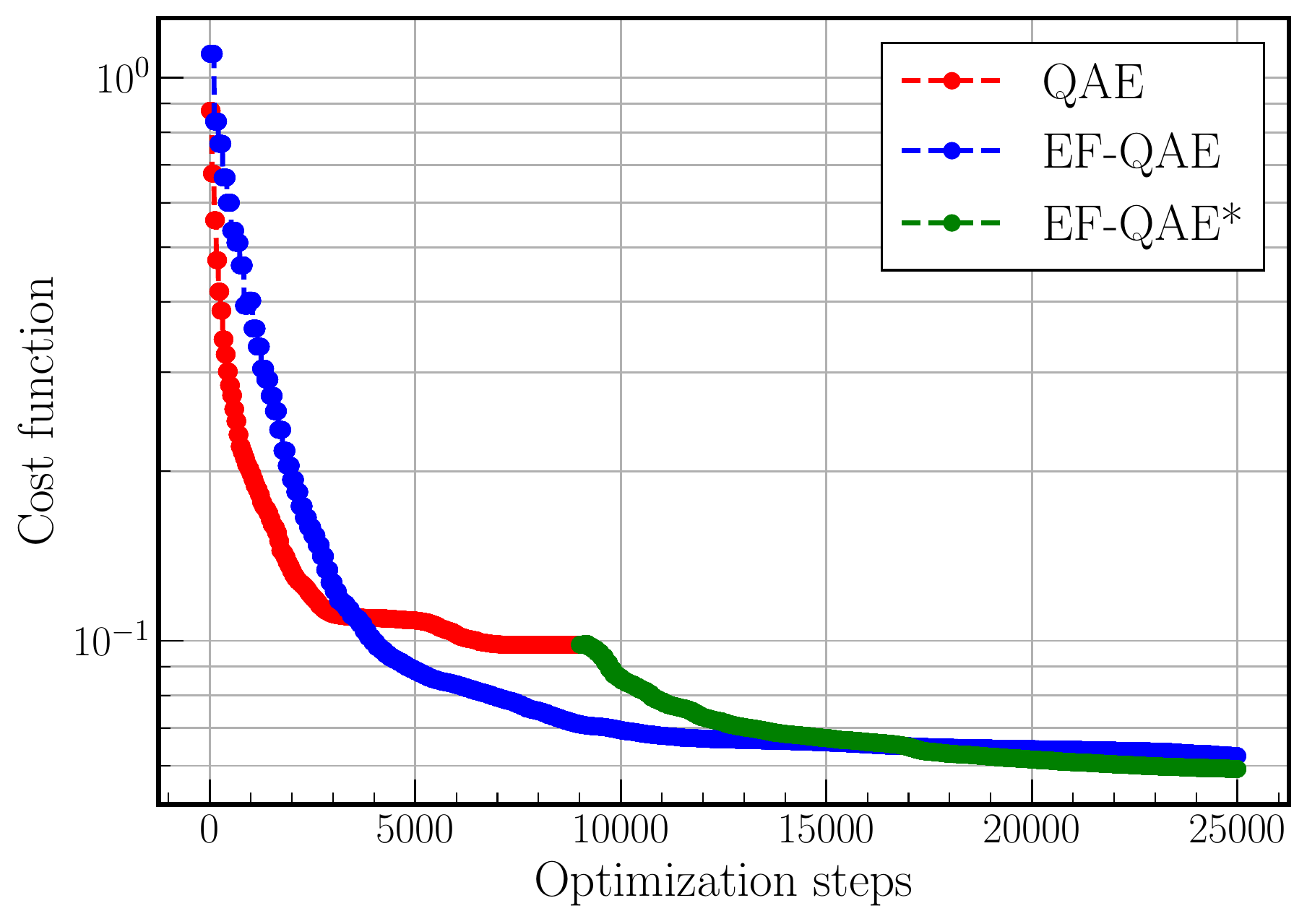}
    \caption{Cost function value as a function of the number of evaluations. Here, we consider the standard QAE, EF-QAE, and EF-QAE*. The EF-QAE* is the EF-QAE initialized with the optimal parameters of QAE. The EF-QAE achieves three times the compression of the QAE using the same quantum resources, at the expense of additional classical optimization.}
    \label{fig:handwritten-steps}
\end{figure}

In Fig. \ref{fig:handwritten-steps}, we show the cost function value as a function of the number of evaluations. Recall that EF-QAE* is simply the EF-QAE initialized with the optimal parameters of QAE. We note that EF-QAE achieves three times the compression of QAE using the same quantum resources. However, in contrast to the previous Ising model case, EF-QAE requires even fewer function evaluations to improve over the standard QAE. This is due to the fact that, although the parameter search space is larger, by including the feature vector we are affecting the parameter landscape in such a way that now it is well-behaved, and therefore, the optimization procedure leads to faster convergence.

\begin{figure}[t]
    \centering
    \includegraphics[width=1.0\linewidth]{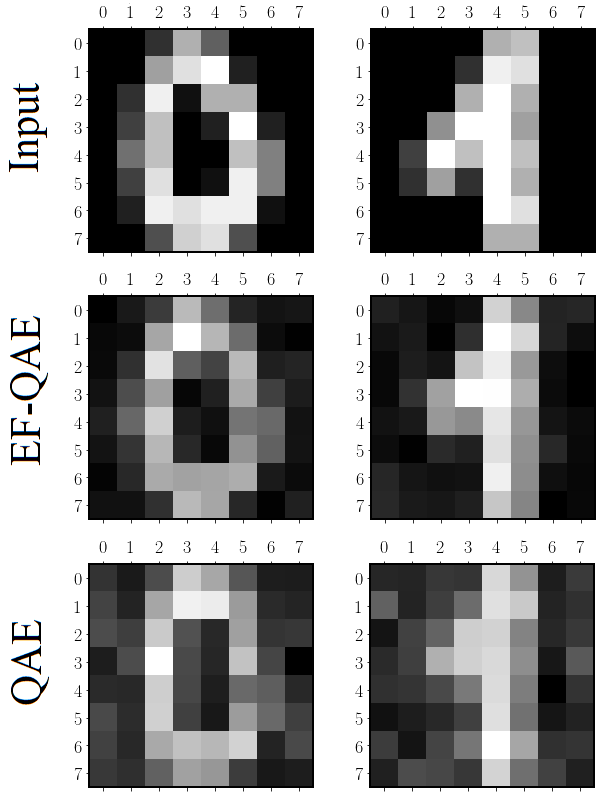}
    \caption{Images of $\mathbf{0}$ and $\mathbf{1}$ handwritten test digits encoded into a 6-qubit state ($8\times 8$ pixels). Images shown correspond to the input state, and the output states of the EF-QAE and QAE models. As can be seen, the fidelity of the EF-QAE output state is improved compared to QAE.}
    \label{fig:handwritten-space}
\end{figure}

Once again, to gain insight into the compression process, we assess both EF-QAE and QAE with the optimal parameters against two handwritten test digits corresponding to $\mathbf{0}$ and $\mathbf{1}$. The results are shown in Fig.~\ref{fig:handwritten-space}. Here, we plot the output digit of the EF-QAE and QAE. Once more, since EF-QAE achieves better compression to the $\ket{00}$ trash state, we obtain higher fidelity on the output state. Remarkably, in both cases, the performance of the EF-QAE is improved with respect to the QAE. In Appendix \ref{sec:fidelities} we discuss and provide the output fidelities of the training and 60 test handwritten digits.

\section{Conclusion} \label{sec:discussion}
We have presented a variational quantum algorithm called EF-QAE capable of compressing quantum data of a parameterized model. In contrast to standard QAE, EF-QAE achieves this compression with higher fidelity. Its key idea is to define a parameterized quantum circuit that depends upon adjustable parameters and a feature vector that characterizes such a model. In this way, the data compression can be tailored to the particular input, informed by the feature vector, and the compression performance is enhanced.

We have validated the EF-QAE in simulations by compressing ground states of the 1D Ising spin chain, and classical handwritten digits encoded into quantum states. We compared the results with the standard QAE. The results show that EF-QAE achieves better compression of the initial state, and therefore, the final output state is recovered with higher fidelity. Moreover, the learning task of EF-QAE can be initialized with the optimal QAE parameters. In this manner, EF-QAE will always improve the QAE performance. Nonetheless, the encoding strategy of the feature vector is amenable to be improved, for instance, allowing the feature vector to be a free variational parameter or using a non-linear encoding. We leave the study of encoding strategies for future work.

The EF-QAE may need additional classical optimization compared to QAE. In contrast, we increase the compression performance using the same amount of limited quantum resources. In this sense, EF-QAE is a step toward what could be done on NISQ computers, shortening the distance between current quantum devices and practical applications. 

\section*{Code availability}
The code is available in \href{https://github.com/Quantum-TII/qibo/tree/master/examples/EF_QAE}{GitHub} \cite{github}.

\section*{Acknowledgements}
The author would like to thank Diego Garc\'ia-Mart\'in and Jos\'e I. Latorre for fruitful discussions. This work is supported by the projects PGC2018-095862-B-C22 and Quantum CAT 001-P-001644.

\bibliographystyle{apsrev4-1}
\bibliography{citations}

\appendix
\section{Comparison table for QAE and EF-QAE} \label{sec:comparison}
In this section we summarize QAE and EF-QAE similarities and differences. The summary is shown in Table \ref{table:summary}.

\begin{table}[h]
\centering\renewcommand\cellalign{c}
\setcellgapes{3pt}\makegapedcells
\begin{tabular}{|c|c|c|} \hline
\textbf{} & \textbf{\,\,\,\,\,\,\,QAE\,\,\,\,\,\,\,} & \textbf{EF-QAE} \\ \hline
\makecell{Quantum resources \\(circuit depth)} & \multicolumn{2}{|c|}{Equal} \\ \hline
Unitary operation & U$(\vec{\theta})$ & U$(\vec{\theta}, \vec{x})$\\ \hline
\makecell{No. trainable parameters \\(in each rotation gate)} & 1 & dim($\vec{x}$) + 1\\ \hline
Classical optimization & \multicolumn{2}{|c|}{\makecell{EF-QAE generally needs \\additional optimization}}\\ \hline
\makecell{Compression \\performance} & \multicolumn{2}{|c|}{\makecell{EF-QAE has always higher \\compression performance}}\\
\hline\end{tabular}
\caption{Summary for QAE and EF-QAE similarities and differences.}
  \label{table:summary}
\end{table}

\section{Output fidelities of test and training sets} \label{sec:fidelities}
In this section, we provide the output fidelities of the training and test sets for the handwritten digit and Ising model examples.

\begin{figure}[h]
    \centering
    \includegraphics[width=1.0\linewidth]{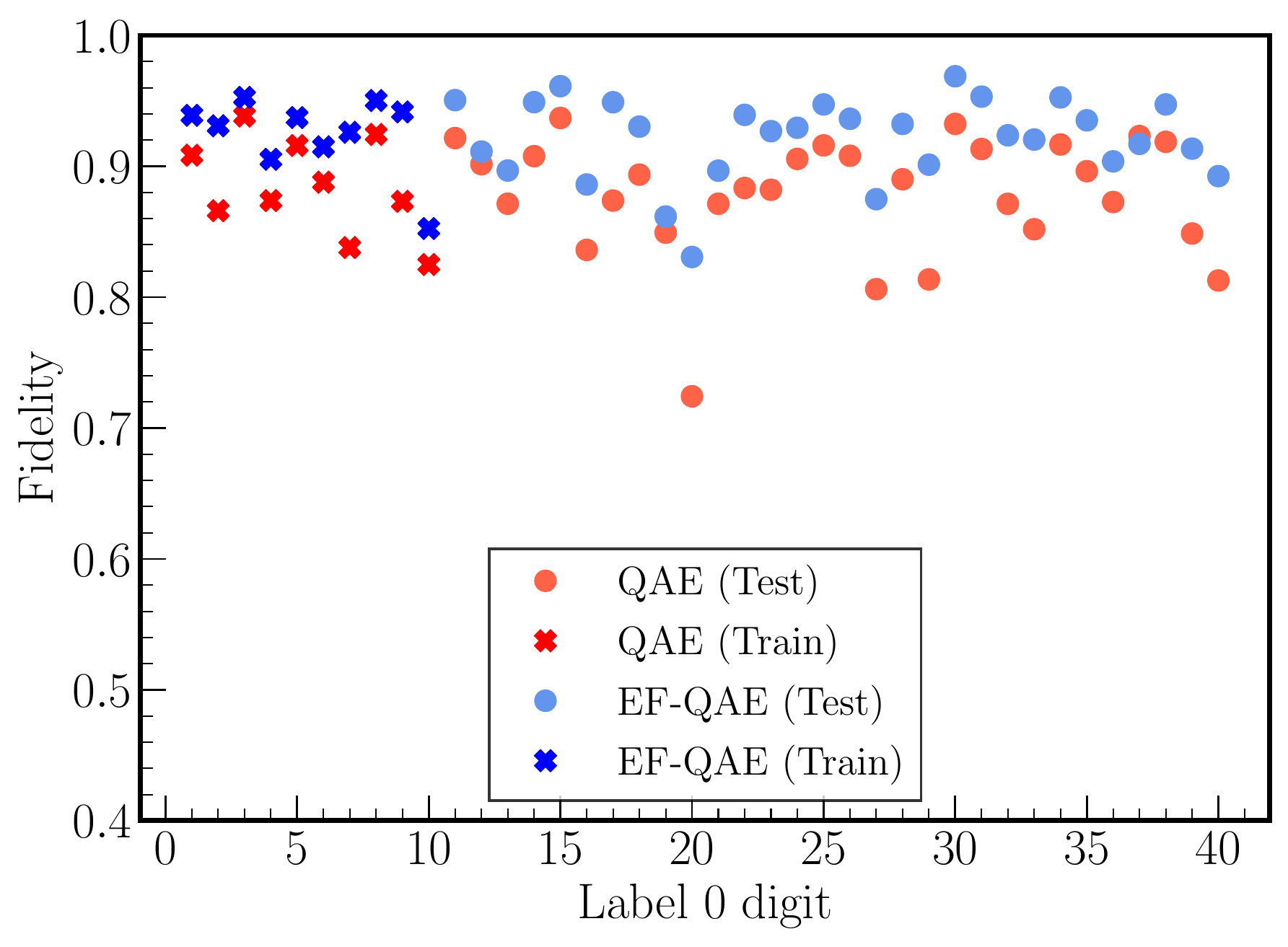}
    \caption{Fidelity of the output state for different \textbf{0} handwritten digits, using the EF-QAE and QAE. We have considered 10 training and 30 test digits.}
    \label{fig:handwritten_fidelities0}
\end{figure}

\begin{figure}[h]
    \centering
    \includegraphics[width=1.0\linewidth]{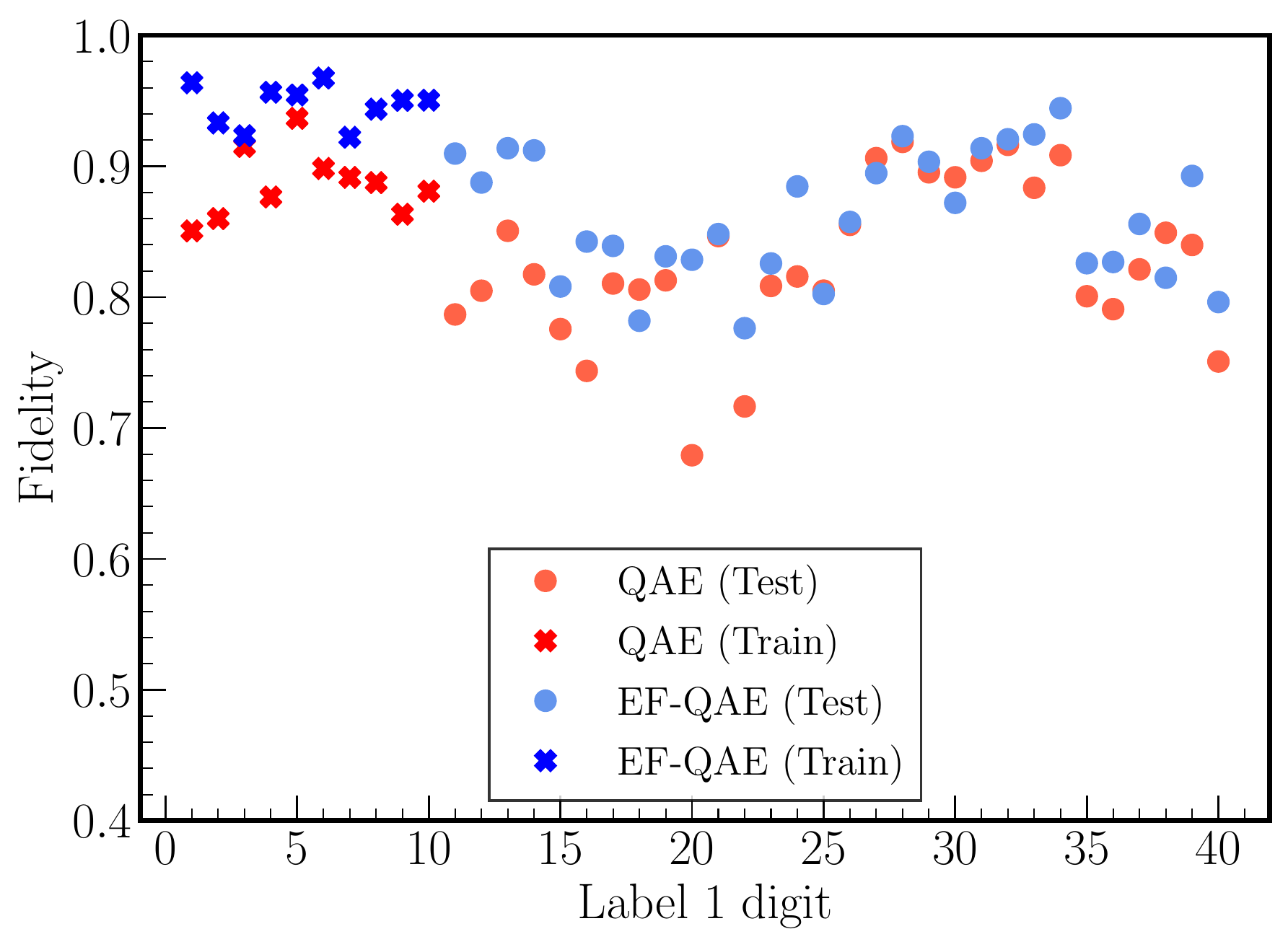}
    \caption{Fidelity of the output state for different \textbf{1} handwritten digits, using the EF-QAE and QAE. We have considered 10 training and 30 test digits.}
    \label{fig:handwritten_fidelities1}
\end{figure}

\textit{Handwritten digits:} In Fig. \ref{fig:handwritten_fidelities0} we show the output fidelities of 10 training and 30 test digits corresponding to the \textbf{0} digit. As can be seen, the performance of the EF-QAE is better, compared to the standard QAE. Similarly, in Fig. \ref{fig:handwritten_fidelities1}, we plot the output fidelities of 10 training and 30 test digits corresponding to the \textbf{1} digit. Here, we observe again that the EF-QAE performance is preferable.

\begin{figure*}[t!]
    \centering
    \includegraphics[width=0.85\linewidth]{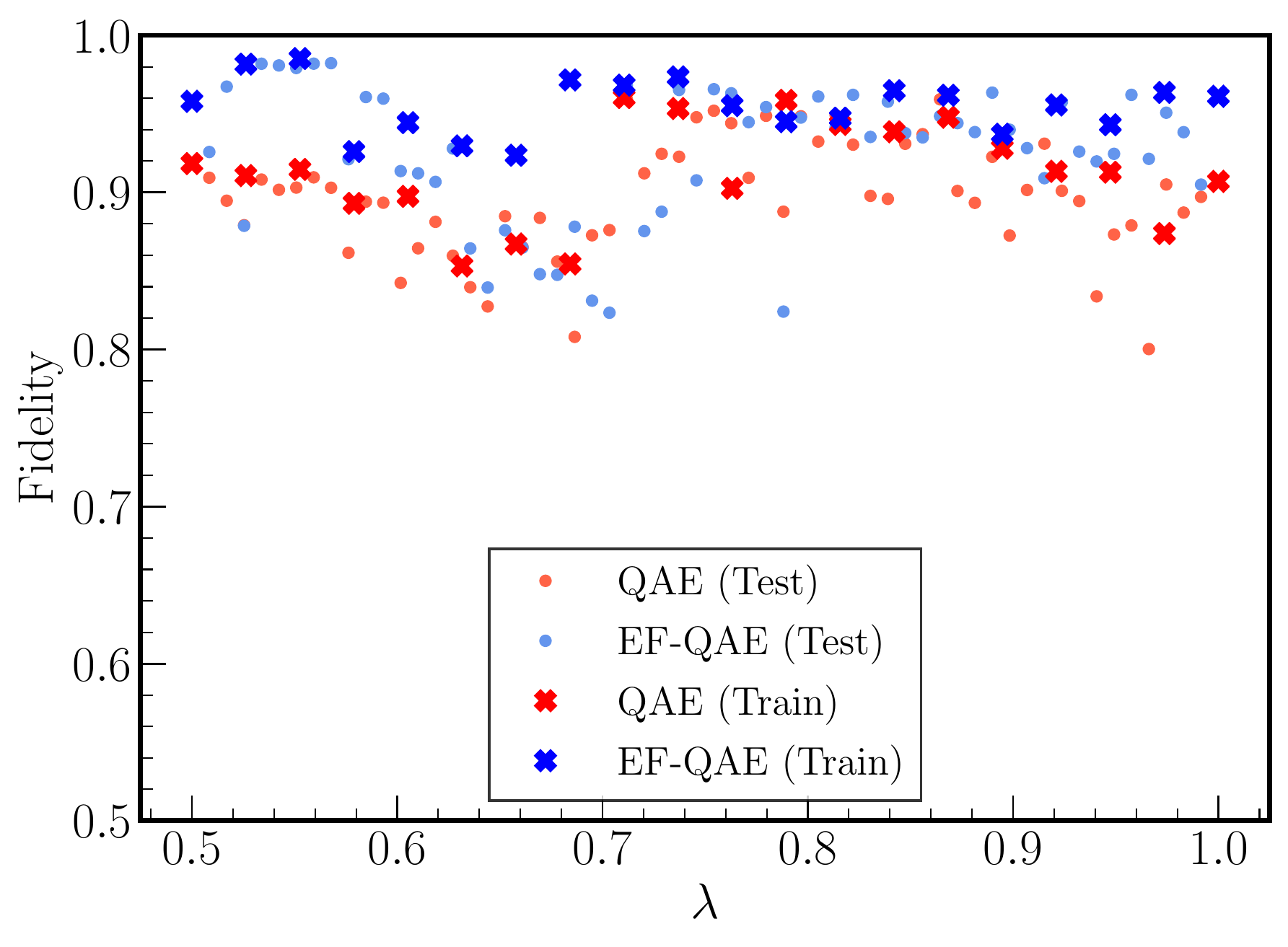}
    \caption{Fidelity of the output state for Ising ground states with different transverse field $\lambda$, using the EF-QAE and QAE. We have considered 20 training and 60 test ground states.}
    \label{fig:ising_fidelities}
\end{figure*}

\textit{Ising model:} In Fig. \ref{fig:ising_fidelities} we show the output fidelities of 20 training and 60 test Ising ground states. As can be seen, the output fidelities of the EF-QAE are higher, except for a few outlier values around $\lambda=0.7$. This could be improved, for instance, by simply increasing the number of training states, or by populating values around $\lambda=0.7$ taking nonequispaced training ground states.

\section{Resilience to noise} \label{sec:resilience_main}
It has been shown recently that specific VQAs can exhibit noise resilience~\cite{sharma2019noise}. That is, the optimal parameters are unaffected by certain noise models. Here we prove that the local cost function $C$ is resilient to global depolarizing noise. Let us rewrite $C$ from Eq. \ref{eq:cost_function} as
\begin{equation}
    C = \frac{1}{2}\sum_{k=1}^{n_t} (1 - \zeta^{(k)})\,,
\end{equation}
where $\zeta^{(k)} = \mte{0}{U^\dagger (Z_k \otimes \id_{\overline{k}} ) U}$. From now on, we refer to $\tilde{C}$ and $\tilde{\zeta}$ as the noisy versions of these quantities. Recall that global depolarizing noise transforms the state according to $\rho \rightarrow q \rho + (1-q) \id /d$. If we consider a circuit that has depth $D$, then the final state is $q^D \rho + (1-q^D) \id /d$.  Notice as well that $\tilde{\zeta}^{(k)}$ is estimated simply by executing the circuit in Fig.~\ref{fig:new_ansatz} and measuring in the computational basis. The maximally mixed state has zero expectation value, since we measure Pauli $Z$ operators. Therefore, we obtain that $\tilde{\zeta}^{(k)} = q^{D}\zeta^{(k)}$, where $D$ is the depth of the circuit used to estimate $\zeta^{(k)}$. This implies
\begin{equation}
    \tilde{C} = \frac{1}{2} \sum_{k=1}^{n_t}(1 -  q^{D} \zeta^{(k)} )\,.
\end{equation}
From this expression, we see that
\begin{equation}
    \argmin_{\vec{\theta}}\tilde{C} = \argmax_{\vec{\theta}}\left(\sum_{k=1}^{n_t} \zeta^{(k)}\right)\,.
\end{equation}
It is clear as well that
\begin{equation}
    \argmin_{\vec{\theta}}C =\argmax_{\vec{\theta}}\left(\sum_{k=1}^{n_t} \zeta^{(k)}\right)\,.
\end{equation}
Hence we arrive at
\begin{equation}
   \argmin_{\vec{\theta}}\tilde{C} = \argmin_{\vec{\theta}}C \,.
\end{equation}
This proves our statement of global depolarizing noise resilience since it shows that the optimal parameters are unaffected.

\end{document}